\documentclass[a4paper]{article}

\usepackage{INTERSPEECH2022}
\usepackage{url}
\usepackage{float}

\title{Voxceleb-ESP: preliminary experiments detecting Spanish celebrities \\from their voices}
\name{Beltrán Labrador, \thanks{The set of video pointer list, segment time stamps in each video, trial lists (A and B) and a python script for automatic downloading and audio segment extraction are freely available under request to \newline audias\texttt{@}uam.es.}
Manuel Otero-Gonzalez, Alicia Lozano-Diez, Daniel Ramos \newline Doroteo T. Toledano, Joaquin Gonzalez-Rodriguez}
\address{
  AUDIAS (Audio, Data Intelligence and Speech), Universidad Autónoma de Madrid, Spain}
\email{\{beltran.labrador, joaquin.gonzalez\}\texttt{@}uam.es}

\begin{document}
\maketitle
\begin{abstract}
This paper presents VoxCeleb-ESP, a collection of pointers and timestamps to YouTube videos facilitating the creation of a novel speaker recognition dataset. VoxCeleb-ESP captures real-world scenarios, incorporating diverse speaking styles, noises, and channel distortions. It includes 160 Spanish celebrities spanning various categories, ensuring a representative distribution across age groups and geographic regions in Spain. We provide two speaker trial lists for speaker identification tasks, each of them with same-video or different-video target trials respectively, accompanied by a cross-lingual evaluation of ResNet pretrained models. Preliminary speaker identification results suggest that the complexity of the detection task in VoxCeleb-ESP is equivalent to that of the original and much larger VoxCeleb in English. VoxCeleb-ESP contributes to the expansion of speaker recognition benchmarks with a comprehensive and diverse dataset for the Spanish language.
\end{abstract}
\noindent\textbf{Index Terms}: speaker recognition, speaker identification, large-scale, dataset

\section{Introduction}

Speaker recognition, the process of identifying or verifying an individual's identity based on their voice, has found wide-ranging applications in various domains. Significant progress in modeling these speaker recognition systems \cite{spk_review, ivectors,td_sid, xvectors} has pushed the accuracy of these models. However, the dataset used for training these models plays a crucial role in determining their performance. The accuracy of these models is highly dependent on the size and quality of the training data, which must comprehensively cover a broad spectrum of factors, including diverse noise environments, channel and microphone distortions, varying speaking styles and accents among other influencing conditions. Moreover, a robust test set is crucial for evaluating model generalization across diverse languages, ensuring effective handling of linguistic variations, global applicability, and enhanced reliability in real-world scenarios.

Several large collections of text-independent speaker recognition data have been developed, including VoxCeleb 1 \cite{voxceleb1}, VoxCeleb 2 \cite{voxceleb2}, Speakers in the Wild \cite{spk_wild}, and CN-Celeb \cite{cn_celeb}. These data collections are freely available for widely used languages like English and Chinese. Additionally, there has been a strong effort to include low-resource languages in these data collections, as shown by the development of a Vietnamese \cite{vietnam_celeb} and an European Portuguese datasets \cite{voxceleb_pt}.
In the Spanish language, some databases have been developed for speaker recognition. The Ahumada corpus \cite{ahumada_icassp,ahumada_speech} is a database designed and acquired to have controlled conditions with the objective of analyzing and measuring the effect of some of the main variability sources found in commercial and forensic applications. On the other hand, SecuVoice \cite{secuvoice} is a speech database that contains sequences of isolated digits from zero to nine acquired using smartphone devices, intended for research on biometrics and secure applications.

In this paper, following the paradigm started by Voxceleb \cite{voxceleb1,voxceleb2}, we introduce Voxceleb-ESP, a dataset for speaker recognition research in Spanish (Castilian) language captured ``in the wild", with the objective of furnishing a comprehensive Spanish dataset. It lacks controlled conditions, offering an increased variability in speaking styles, noise, and channel distortions. We obtained the audio from YouTube videos and meticulously manually curated the database, leveraging elements from the automated pipeline proposed in \cite{voxceleb1}. The database comprises around 7 hours of voice from 160 Spanish celebrities, spanning diverse categories. It exhibits extensive social and regional variability, representing various geographic zones of Spain, a broad age range and a balanced gender distribution.

In order to establish a initial benchmark to evaluate the performance on the speaker identification task, we have created two different speaker trial lists with different conditions and variability. Finally, we have conducted diverse experiments evaluating various systems pretrained on Voxceleb 2 \cite{voxceleb2} for their performance on this Voxceleb-ESP collection.

The rest of this paper is organized as follows. Section \ref{sec:database} details the database description and trial lists, Section \ref{sec:experiments} describes the experiments, and we conclude in Section \ref{sec:conclusion}.

\section{Database description}
\label{sec:database}

To create this database of Spanish celebrities, our initial step involved curating a diverse list of persons of interest. We aimed to encompass a broad spectrum of Spanish public figures, including singers, journalists, television hosts, actors, politicians, athletes and comedians, with a distribution detailed in Table \ref{table:categories}. A crucial consideration was maintaining gender balance, resulting in a proportional representation of 51.25\% male and 48.75\% female across categories. Additionally, we prioritized social and regional diversity, incorporating celebrities from various geographic regions of Spain. To enhance the database's variability, we selected celebrities spanning ages from 20 to 70 years old. The celebrities selected are listed in the Annex.

\begin{table}[ht!]
\centering
\caption{Categories distribution of celebrities}
\begin{tabular}{lccc}
\toprule
\textbf{Categories} & \textbf{Total} & \textbf{Male} & \textbf{Female} \\
\hline
Singers & 25 & 13 & 12 \\
Journalists & 12 & 6 & 6 \\
TV Hosts & 26 & 14 & 12 \\
Actors & 24 & 12 & 12 \\
Politicians & 31 & 15 & 16 \\
Athletes & 23 & 12 & 11 \\
Comedians & 19 & 10 & 9 \\
\hline
\textbf{Total} & 160 & 82 & 78 \\
\bottomrule
\end{tabular}
\label{table:categories}
\end{table}

Each selected speaker contributes to the database with the audio from three YouTube videos, thus featuring speech in three distinct acoustic conditions. While selecting the videos we aim to introduce various extrinsic factors in each video, such as different acoustic settings (television studios, radio booths, rooms, outdoor environments), diverse audio capture systems (common microphones in devices like computers or mobile phones or professional microphones in television and radio studios), and background sounds (laughter, shouts, music, ambient noise), enhancing the variability of the dataset.

Within each video audio, five segments containing only the celebrity's speech are extracted manually. Thus, each speaker yields 15 audio segments, resulting in a balanced database of 2,400 fragments, with 1,245 from male speakers and 1,155 from female speakers. The segments, approximately 10 seconds in duration (ranging from 8 to 12 seconds), collectively amount to over six and a half hours of audio, as detailed in Table \ref{table:distribution}.

\begin{table}[ht]
\centering
\caption{Distribution of 10s segments and total speech time}
\begin{tabular}{lcccc}
\toprule
\textbf{Distribution} & \textbf{Speakers} & \textbf{Videos} & \textbf{Segments} & \textbf{Time} \\
\hline
Total & 160 & 480 & 2,400 & 6h 40min \\
Male & 83 & 249 & 1,245 & 3h 28min \\
Female & 77 & 231 & 1,155 & 3h 12min \\
\bottomrule
\end{tabular}
\label{table:distribution}
\end{table}

The audio from each YouTube video is automatically downloaded using the Youtube-dl tool in \textit{wav} format. FFmpeg \cite{ffmpeg} is then employed to resample to 16000 samples per second, extract a single channel, and trim the selected five segments.

\subsection{Speaker identification trial lists}

We have crafted two distinct trial lists. Trial List A \textit{target} trials consists segments of the same speaker within a single video, simplifying speaker identification due to the smaller variability. \textit{Non-target} trials involve different speakers from separate videos.

Trial List B pairs each speaker \textit{target} trials with segments from different videos, introducing more variability and posing a greater challenge to the speaker identification. \textit{Non-target} trials are created using different speakers from distinct videos. Both evaluation trial lists are further divided into male and female trials.
\begin{table}[ht]
    \centering
    \caption{Number of trials in list A (same-video target trials) and list B (different-video target trials)}
    \begin{tabular}{l|ccc}
    \toprule
        & \multicolumn{3}{c}{\textbf{Trial List A}} \\
        & \# Trials & \# Target & \# Non-target \\
        \hline
        Male & 10,824 & 984 & 9,840 \\
        Female & 10,296 & 936 & 9,360 \\
        Total & 21,120 & 1,920 & 19,200 \\
        \midrule
        & \multicolumn{3}{c}{\textbf{Trial List B}} \\
        & \# Trials & \# Target & \# Non-target \\
        \hline
        Male & 45,100 & 4,100 & 41,000 \\
        Female & 42,900 & 3,900 & 39,000 \\
        Total & 88,000 & 8,000 & 80,000 \\
        \bottomrule
    \end{tabular}
    \label{your_label}
\end{table}

\section{Experiments and results}
\label{sec:experiments}

To establish an initial benchmark on this database, we assessed the performance of two state-of-the-art models pretrained on the Voxceleb 2 \cite{voxceleb2} English dataset. We employed the convolutional models ResNetSE34L, detailed in \cite{voxceleb_trainer}, and ResNetSE34v2, described in \cite{voxceleb_trainer2}. The models weights and code are freely accessible at \url{https://github.com/clovaai/voxceleb_trainer}.

\begin{table}[ht]
  \centering
  \caption{Performance in Equal Error Rate (EER) of the English-only pretrained models on different speaker identification evaluation tasks with spanish-only trials}
  \begin{tabular}{l|ccc}
    \toprule
      \textbf{Model} & \textbf{ResNetSE34L} & \textbf{ResNetSE34v2} \\ 
      & \multicolumn{2}{c}{VoxCeleb 1} \\ \hline
      All & 2.17\% & 1.17\% \\ 
      Male & 3.03\% & 1.67\% \\
      Female & 3.47\% & 1.3\% \\
      & \multicolumn{2}{c}{VoxCeleb-ESP - Trial List A} \\ \hline
      All & 1.97\% & 1.51\% \\ 
      Male & 1.51\% & 1.01\% \\
      Female & 3.64\% & 2.63\% \\
      & \multicolumn{2}{c}{VoxCeleb-ESP - Trial List B} \\ \hline
      All & 5.46\% & 3.15\% \\
      Male & 4.63\% & 2.43\% \\
      Female & 7.94\% & 4.79\% \\
      \bottomrule
  \end{tabular}
  \label{table:results}
\end{table}

Table \ref{table:results}, illustrate the results in Equal Error Rate (EER) that we obtained on both the VoxCeleb 1 test set and on the presented Voxceleb-ESP trial lists. We show the results on the male and female only trials on the three trial lists, excluding any cross-gender trial.
These results demonstrate the performance of the systems solely trained on the English language database Voxceleb 2, when applied to the presented Spanish task without any adaptation. These models showcase a good ability to generalize to an unseen language. Nevertheless, adaptation strategies as Probabilistic Linear Discriminant Analysis (PLDA) or incorporating specific Spanish data could enhance the effectiveness in this particular scenario. 
The systems show better performance in Trial List A compared to Trial List B, which indicates that Trial List A represents a comparatively less complex task, attributed to the reduced variability of the \textit{target} speaker trials segments, as these trial pairs are extracted from the same video, with the same acoustic and environmental conditions.

\section{Conclusions}
\label{sec:conclusion}
In this paper, we present VoxCeleb-ESP, a Spanish dataset for speaker recognition. As a language-specific extention to Voceleb, it features audio extracted from YouTube videos from Spanish speaking celebrities. Two trial list with varying conditions and difficulty are provided for speaker identification tasks. Additionally, an initial performance benchmark is established through cross-lingual evaluation of models trained on the English VoxCeleb 2 database. Voxceleb-ESP contributes to expanding the speaker identification evaluation benchmarks and fosters for further research in multilingual speaker recognition.

\section{Acknowledgements}
 This work has been supported by the FPI RTI2018-098091-B-I00,  MCIU/AEI/10.13039/501100011033/FEDER, UE and \\PID2021-125943OB-I00,  MCIN/AEI/10.13039/501100011033/\\FEDER, UE from the Spanish Ministerio de Ciencia e Innovaci\'on and Fondo Europeo de Desarrollo Regional.
\bibliographystyle{IEEEtran}

\bibliography{mybib}


\newpage
\onecolumn
\section{Annex}
\label{sec:annex}
\begin{table}[H]
    \centering
    \caption{Spanish celebrities selected for building VoxCeleb-ESP}
    \begin{tabular}{c c c c}
        \toprule
        \multicolumn{2}{c}{\textbf{Singers}} & \multicolumn{2}{c}{\textbf{Comedians}} \\
        \textbf{Male} & \textbf{Female} & \textbf{Male} & \textbf{Female} \\
        \hline
        Alejandro Sanz & Aitana Ocaña & Alex Clavero & Ana Morgade \\
        Antonio Orozco & Ana Belén & Berto Romero & Anabel Alonso \\
        Beret & Ana Mena & Dani Rovira & Carolina Iglesias \\
        C Tangana & India Martínez & Ernesto Sevilla & Eva Hache \\
        Dani Martín & Isabel Pantoja & Goyo Jiménez & Henar Álvarez \\
        David Bisbal & Leire Martínez & Joaquín Reyes & Silvia Abril \\
        Joaquín Sabina & Lola Índigo & José Mota & Valeria Ros \\
        Manuel Carrasco & Malú & Leo Harlem & Victoria Martin \\
        Melendi & Rocío Jurado & Miguel Gila & Yolanda Ramos \\
        Omar Montes & Rosalía Vila & Miguel Lago & \\
        Pablo Alborán & Rosario Flores & & \\
        Pablo López & Rozalén & & \\
        Raphael & & & \\

        \multicolumn{2}{c}{\textbf{Television Hosts}} & \multicolumn{2}{c}{\textbf{Actors}} \\
        \textbf{Male} & \textbf{Female} & \textbf{Male} & \textbf{Female} \\
        \hline
        Arturo Valls & Ana Obregón & Alejandro Amenábar & Anna Castillo \\
        Carlos Sobera & Ana Rosa Quintana & Antonio Banderas & Blanca Suárez \\
        Dani Martínez & Anne Igartiburu & Antonio Resines & Ester Espósito \\
        David Broncano & Cristina Pedroche & Javier Bardem & Inma Cuesta \\
        Frank Blanco & Irene Junquera & Mario Casas & Maribel Verdú \\
        Gran Wyoming & Mercedes Milá & Miguel Ángel Silvestre & Paz Vega \\
        Iker Jiménez & Nuria Roca & Alex de la Iglesia & Belén Rueda \\
        Jesús Vázquez & Paz Padilla & Antonio Garrido & Elsa Pataky\\
        Jordi Hurtado & Pilar Rubio & Javier Cámara & Isabel Coixet \\
        Jorge Javier Vázquez & Sandra Golpe & José Coronado & Macarena García \\
        Juan y Medio & Sandra Sabatés & Pedro Almodóvar & Penélope Cruz \\
        Luis Larrodera & Sonsoles Ónega & Santiago Segura & Úrsula Corberó \\
        Manel Fuentes & & & \\
        Pablo Motos & & & \\

        \multicolumn{2}{c}{\textbf{Politicians}} & \multicolumn{2}{c}{\textbf{Athletes}} \\
        \textbf{Male} & \textbf{Female} & \textbf{Male} & \textbf{Female} \\
        \hline
        Albert Rivera & Ada Colau & Alejandro Valverde & Adriana Cerezo \\
        Felipe González & Cayetana Álvarez de Toledo & Andrés Iniesta & Alexia Putellas \\
        Juan Manuel Moreno & Irene Montero & Javier Fernández López & Ana Carrasco \\
        Mariano Rajoy & Isabel Díaz Ayuso & Joaquin Sanchez & Ana Peleteiro \\
        Pablo Casado & Manuela Carmena & Pau Gasol & Mireia Belmonte \\
        Martínez Almeida & Macarena Olona & Fernando Alonso & Almudena Cid \\
        Miguel Ángel Revilla & María Carmen Calvo & Iker Casillas & Anaya Valdemoro \\
        Núñez Feijoo & María Jesús Montero & John Rahm & Carolina Marín \\
        Pablo Iglesias & Margarita Robles & Juan Carlos Navarro & Garbiñe Muguruza \\
        Pedro Sánchez & Meritxell Batet & Marc Márquez & Lydia Valentín \\ 
        Rubalcaba & Nadia Calviño & Rafael Nadal & Ona Carbonell \\
        Santiago Abascal & Rocío Monasterio & Raúl González & \\
        Jose Luis Rodríguez Zapatero & Soraya Sáenz & & \\
        Íñigo Errejón & Cristina Cifuentes & & \\
        Jose María Aznar & Inés Arrimadas & & \\
        & Yolanda Díaz & & \\

        \multicolumn{2}{c}{\textbf{Journalists}} &  \\
        \textbf{Male} & \textbf{Female} &  & \\
        \hline
        Antonio García Ferreras & Ana Pastor & & \\
        Carlos Alsina & Cristina Pardo & & \\
        Carlos Herrera & Helena Resano & & \\
        Josep Pedrerol & Julia Otero & & \\
        Matías Prats & Mónica Carrillo & & \\
        Pedro Piqueras & Susanna Griso & & \\
        \bottomrule
    \end{tabular}
    \label{tab:spanish_celebrities}
\end{table}

\end{document}